\documentclass[aps, twocolumn, letterpaper]{revtex4}

\usepackage{amsmath}
\usepackage{amssymb}
\usepackage{mathrsfs}
\usepackage{xspace}
\usepackage{graphicx}
\usepackage{braket}
\usepackage[utf8]{inputenc}

\usepackage{ifpdf}
\ifpdf
\fi

\newcommand{\eq}[1]{(\ref{#1})}
\newcommand{\Eq}[1]{Eq.~(\ref{#1})}
\newcommand{\Eqs}[1]{Eqs.~(\ref{#1})}

\newcommand{\Sec}[1]{Sec.~\ref{#1}}

\newcommand{\Ref}[1]{Ref.~\cite{#1}}
\newcommand{\Refs}[1]{Refs.~\cite{#1}}

\newcommand{\eg}{{e.g.,\/}\xspace}				
\newcommand{\ie}{{i.e.,\/}\xspace}				

\newcommand{\pd}{\partial}
\newcommand{\del}{\vec{\nabla}}
\newcommand{\cc}{\text{c.\,c.}}
\newcommand{\hc}{\text{h.\,c.}}

\newcommand{\mc}[1]{\mathcal{#1}}
\newcommand{\mcc}[1]{\mathfrak{#1}}

\newcommand{\oper}[1]{\hat{\vec{#1}}}
\renewcommand{\vec}[1]{{\boldsymbol{\rm #1}}}

\usepackage{color}

\begin{document}

\title{First-principle variational formulation of polarization effects in geometrical optics}

\author{D.~E. Ruiz and I.~Y. Dodin}
\affiliation{Department of Astrophysical Sciences, Princeton University, Princeton, New Jersey 08544, USA}
\date{\today}

\begin{abstract}

The propagation of electromagnetic waves in isotropic dielectric media with local dispersion is studied under the assumption of small but nonvanishing $\lambda/\ell$, where $\lambda$ is the wavelength, and $\ell$ is the characteristic inhomogeneity scale. It is commonly known that, due to nonzero $\lambda/\ell$, such waves can experience polarization-driven bending of ray trajectories and polarization dynamics that can be interpreted as the precession of the wave ``spin''. The present work reports how Lagrangians describing these effects can be deduced, rather than guessed, within a strictly classical theory. In addition to the commonly known ray Lagrangian that features the Berry connection, a simple alternative Lagrangian is proposed that naturally has a canonical form. The presented theory captures not only the eigenray dynamics but also the dynamics of continuous wave fields and rays with mixed polarization, or ``entangled'' waves. The calculation assumes stationary lossless media with isotropic local dispersion, but generalizations to other media are straightforward to do.

\end{abstract}

\maketitle

\bibliographystyle{apsrev-title}

\section{Introduction}

Electromagnetic (EM) waves propagating in inhomogeneous linear media exhibit a variety of intriguing phenomena associated with wave polarization. Those include precession of the polarization vector, known as the Rytov rotation \cite{Rytov:1938wp,Vladimirskii:1941wa,Tomita:1986jo,Berry:1987cj}, and also the polarization-driven bending of ray trajectories, known as the optical Magnus effect or the optical Hall effect \cite{Dugin:1991tx,Dooghin:1992kr,Liberman:1992bz,Onoda:2004ij}. As overviewed recently in \Ref{Bliokh:2015aa}, these phenomena can be attributed as manifestations of the Berry phase \cite{Berry:1984jv}, which is a fundamental concept emerging also in many other areas of physics \cite{Wilczek:1989wn,Bohm:2003wq,Murakami:2003by}. Hence the subject has been attracting an increased amount of attention, especially with focus on variational formulations, which are particularly elucidating at studying EM polarization effects in the geometrical-optics (GO) limit \cite{Liberman:1992bz,Bliokh:2004ja,Bliokh:2004kp,Onoda:2006gg,foot:berk}. Yet a straightforward universal theory is still lacking. Within existing models, ray Lagrangians and Hamiltonians either have to be guessed or are derived using additional postulates like quantization, which is, by definition, alien to classical theory. Such \textit{ad~hoc} approaches result in a number of limitations; \eg they are not readily applicable to media with nonlocal and strongly anisotropic dispersion like magnetized plasma. Thus, there is a compelling need for a generalized and simplified description of EM polarization effects from first principles.

The purpose of this paper is to present such a description. More specifically, what we report here is a new application of the general theory that we derived earlier in  \Ref{Ruiz:2015vq} for the Dirac electron. In the present paper, we explain how \textit{exactly the same theory} can be applied to GO EM waves and polarization effects in particular. As opposed to \Ref{Onoda:2006gg}, our derivation is entirely classical. It is also straightforward and elementary in the sense that the wave Lagrangian does not need to be guessed, as in \Refs{Liberman:1992bz,Bliokh:2008km}, but is rather deduced (basically, using nothing more than matrix multiplication) according to a formalized algorithm. The known results are successfully reproduced and are extended as follows: (i)~We show that, in addition to the commonly known ray Lagrangian that features the Berry connection \cite{Bliokh:2005be,Bliokh:2007fr,Bliokh:2008km}, an alternative ray Lagrangian is possible that naturally has a canonical form and corresponds to a Hamiltonian simpler than that proposed in \Ref{Liberman:1992bz}. We explain how the two Lagrangians are related and demonstrate their (approximate) equivalence numerically. (ii)~As opposed to the result of \Ref{Bliokh:2008km}, our Lagrangians are expressed in terms of physical time and wave vector, so they capture the complete ray dynamics rather than just the ray trajectory. (iii)~In addition to eigenray equations, we also derive equations for continuous wave fields and rays with mixed polarization, or ``entangled'' waves. This description captures both the Rytov rotation and the optical Hall effect simultaneously, and it is also manifestly conservative, since the amplitude equations too are derived from a variational principle.

The calculation presented below applies to arbitrary linear stationary lossless media with isotropic local dispersion, \ie media whose dielectric and magnetic permittivities are real scalars depending only on spatial coordinates. However, within our theory, generalizations to strongly anisotropic media with nonlocal dispersion, such as magnetized plasma, are straightforward to do. The present paper is intended as an introduction to such calculations, which will be reported separately. Thus, below we intentionally focus on a relatively simple problem to show how our general theory fits existing literature.

This work is organized as follows. In \Sec{sec:notation}, we define the basic notation. In \Sec{sec:basic}, we present a Lagrangian formalism to describe EM wave propagation in isotropic stationary dielectric media with local dispersion. In \Sec{sec:reduced}, we obtain a reduced model which captures first-order polarization effects in transverse EM waves. In \Sec{sec:wave}, we discuss a fluid Lagrangian model, which describes the dynamics of the wave envelope, ray trajectory, and polarization. In \Sec{sec:ray_can}, we derive ray equations in canonical form. In \Sec{sec:ray_noncan}, the noncanonical ray equations are presented and compared with those previously reported. In \Sec{sec:conclusion}, we summarize our main results.

\section{Notation}
\label{sec:notation}

The following notation is used throughout this work. The symbol ``$\doteq$'' denotes definitions, ``\hc'' denotes ``Hermitian conjugate'', and ``\cc '' denotes ``complex conjugate''. Also, $\mathbb{I}_N$ denotes a unit $N \times N$ matrix, and hat ($\hat{\hphantom{k}}$) is reserved for differential operators. The Minkowski metric is adopted with signature $(+, -, -, -)$, so, in particular, $\mathrm{d}^4x \equiv \mathrm{d}t\,\mathrm{d}^3x$. Generalizations to curved metrics are straightforward to apply \cite{Dodin:2014hw}. Greek indices span from $0$ to $3$ and refer to spacetime coordinates, $x^\mu$, with $x^0$ corresponding to the time variable, $t$; in particular, $\pd_\mu \equiv \pd/\pd x^\mu$. Latin indices span from $1$ to $3$ and denote the spatial variables, $x^i$ (except where specified otherwise); in particular, $\pd_i \equiv \pd/\pd x^i$. Summation over repeated spatial indices is assumed. Also, in Euler-Lagrange equations (ELEs), the denotation ``$\delta a: $'' means, as usual, that the corresponding equation was obtained by varying the action integral with respect to $a$.

\section{Basic equations}
\label{sec:basic}

\subsection{Photon wave function}

In this paper, we consider EM wave propagation in isotropic lossless dielectric media with local linear dispersion. In this case, the governing equations for the electric field $\vec{E}(t,\vec{x})$ and magnetic field $\vec{H}(t,\vec{x})$ are 
\begin{gather}
	\pd_t \vec{E} 	=  		\frac{c}{\varepsilon} 	\del \times \vec{H},			\label{eq:ampere}\\
	\pd_t \vec{H}		= 	- 	\frac{c}{\mu}				\del \times \vec{E},			\label{eq:faraday}
\end{gather}
where $\varepsilon=\varepsilon(\vec{x})$ is the electric permittivity, and $\mu=\mu(\vec{x})$ is the magnetic permeability. 
 
Let us introduce the following normalized fields,
%
\begin{gather}\label{eq:new_var}
 	\bar{\vec{E}} 		=		\sqrt{\varepsilon}	 \, \vec{E} 	, \\
	\bar{\vec{H}} 		=		\sqrt{\mu}	 \, 		\vec{H} 	,		            
\end{gather}
so one can rewrite \Eqs{eq:ampere} and \eq{eq:faraday} as
\begin{gather}
\pd_t \bar{\vec{E}} =    \frac{c}{n} \del \times \bar{\vec{H}}  -\frac{c}{n} \del \left( \ln \sqrt{\mu}			\right) \times		\bar{\vec{H}}	,		\label{eq:ampere_II}\\
\pd_t \bar{\vec{H}} =  -\frac{c}{n} \del \times \bar{\vec{E}}  +\frac{c}{n} \del \left( \ln \sqrt{\varepsilon}	\right) \times		\bar{\vec{E}} 	,		\label{eq:faraday_II} 
\end{gather}
where $n(\vec{x}) \doteq  \sqrt{\varepsilon \mu}$ is the refraction index. Due to the linearity of these equations, we may formally extend them to complex fields. Then, we can express \Eqs{eq:ampere_II} and \eq{eq:faraday_II} as a vector Schrödinger equation:
\begin{gather}\label{eq:schrodinger}
	i\pd_t \psi = H(\vec{x},-i \del ) \psi,
\end{gather}
where
\begin{equation}\label{eq:psi}
	\psi(t,\vec{x}) \doteq \begin{pmatrix}  \bar{\vec{E}} \\ \bar{\vec{H}}  \end{pmatrix}
\end{equation}
is a six-component wave function, which can be interpreted as the photon wave function \cite{BialynickiBirula:1996wn,Dodin:2014hw}. The Hamiltonian operator is a $6\times 6$ matrix given by
\begin{equation}\label{eq:hamiltonian}
	H(\vec{x}, \oper{k} ) \doteq \frac{c}{n} \vec{\lambda} \cdot \oper{k} +  \mc{A}  ,
\end{equation}
where $\oper{k} \doteq -i \del$ is the wave vector (momentum) operator, the $\vec{\lambda}$ matrices are $6\times 6$ Hermitian matrices,
\begin{equation}\label{eq:gamma}
\vec{\lambda} \doteq
	\begin{pmatrix}
		0 & i\vec{\alpha} \\
		-i \vec{\alpha} & 0
		\end{pmatrix},
\end{equation}
and the matrix $\mc{A}(\vec{x})$ is defined as follows:
\begin{equation}
\mc{A}(\vec{x})  \doteq		\frac{c}{n}
	 \begin{pmatrix}  0 & 	-\vec{\alpha} \cdot \del (\ln \sqrt{\mu})	 \\  \vec{\alpha} \cdot \del (\ln \sqrt{\varepsilon}) 	&  0    \end{pmatrix}   .
\end{equation}
Here, the $\vec{\alpha}$ matrices are $3\times 3$ traceless Hermitian matrices given by \cite{foot:GellMann}
\begin{gather} \label{eq:alpha}
	\alpha^x \doteq \begin{pmatrix}
		0 & 0 & 0 \\
		0 & 0 & -i \\
		0 & i & 0 
	\end{pmatrix}, \\
	\alpha^y \doteq 
	\begin{pmatrix}
		0 & 0 & i \\
		0 & 0 & 0 \\
		-i & 0 & 0 
	\end{pmatrix}, \\
	\alpha^z \doteq
	\begin{pmatrix}
		0 & -i & 0 \\
		i & 0 & 0 \\
		0 & 0 & 0 
	\end{pmatrix}.
\end{gather}
These matrices serve as generators for the vector product. Namely, for any two column vectors $\vec{A}$ and $\vec{B}$, one has
\begin{gather}
 ( \vec{\alpha} \cdot \vec{A}) \vec{B}= i \vec{A} \times \vec{B}, 		\label{eq:prop_I} 		\\
 \vec{A}^T \alpha^j \vec{B}= -i (\vec{A}  \times \vec{B} )^j ,			\label{eq:prop_II}
\end{gather}
where $T$ denotes the matrix transpose.

Let us also point out that, since
%
the Hamiltonian is Hermitian, it conserves the wave action: $\pd_\mu \left( \psi^\dag \gamma^\mu 	\psi \right) =0$,
where $\pd_\mu=(\pd_t ,  \del )$, and $\gamma^\mu \doteq ( \mathbb{I}_6,  c \vec{\lambda} / n )$. More explicitly, this can be written as
\begin{equation}
\frac{\pd}{\pd t}\left( \frac{|\bar{\vec{E}}|^2+|\bar{\vec{H}}|^2}{16 \pi} 	\right) 
	+ 	\del \cdot \left[ \frac{c}{n}		\frac{ \mathrm{Re}  (\bar{\vec{E}}^* \times \bar{\vec{H}} ) }{8\pi}		\right]=0,
\end{equation}
where the $8\pi$ factor was introduced to emphasize the connection to the well known Poynting theorem \cite{Jackson:1999uq}.

\subsection{Lagrangian density}

Equation \eq{eq:schrodinger} has a form akin to the so-called multisymplectic form \cite{Bridges:2001aa,Bridges:2006aa,Bridges:2006bb} and can be readily given a variational interpretation. Specifically, \Eq{eq:schrodinger} can be obtained as the Euler-Lagrange equation, $\delta \Lambda = 0$, for the action integral
\begin{gather}\label{eq:actint}
\Lambda = \int \mcc{L}	\,	\mathrm{d}^4x.
\end{gather}
where the Lagrangian density $\mcc{L}$ is given by
\begin{equation}\label{eq:lagr}
\mcc{L} = \frac{i}{2} \left[ \psi^\dag \gamma^\mu (\pd_\mu \psi) -  (\pd_\mu \psi^\dag) \gamma^\mu \psi   \right] +  \psi^\dag \mc{M} \psi, 
\end{equation}
Here we adopted natural units, such that $c = 1$. We also introduced
\begin{equation}
\mc{M} (\vec{x}) =	 \frac{1}{2n}	\begin{pmatrix}  0 & \vec{\alpha} \\  \vec{\alpha}  &  0    \end{pmatrix} \cdot  \del	\ln	 Z(\vec{x}) ,
\end{equation}
where $Z(\vec{x})$ is the impedance of the medium,
\begin{equation}
Z(\vec{x})	\doteq	\sqrt{\frac{\mu}{\varepsilon}}.
\end{equation}
It is to be noted that no approximations have been used in order to obtain this Lagrangian model. Notice also that the Lagrangian density \eq{eq:lagr} involves only first-order differential operators. In \Ref{Ruiz:2015vq}, we showed that such a simple structure is convenient for studying a point-particle model of the spin-orbit coupling for the Dirac electron. Below, we report how the calculation can also be extended to study the effect of polarization of classical waves in dielectrics of the specified type.

\section{Reduced Model}
\label{sec:reduced}

We consider waves such that
\begin{equation}
\psi(t,\vec{x})=\xi e^{i\theta},
\end{equation}
where $\theta(t,\vec{x})$ is some rapid real phase, and $\xi(t,\vec{x})$ is a vector evolving slowly compared to $\theta(t,\vec{x})$. Hence, we assume that the inhomogeneity scale $\ell$ is large compared to the wavelength; \ie

\begin{equation}
\epsilon\doteq \frac{1}{k  \ell} \ll 1,
\label{eq:ordering}
\end{equation}
where $k\doteq|\vec{k}|$, and 
\begin{equation}
\vec{k}\doteq \del \theta
\end{equation}
is the wave vector. Below, we construct a reduced model of EM wave propagation using the smallness of $\epsilon$. Unlike the standard geometrical-optics (GO) theory, which corresponds to a Lagrangian accurate to the zeroth order in $\epsilon$ \cite{Tracy:2014to,Dodin:2012hn}, this reduced model will yield a ray Lagrangian accurate to the first order in $\epsilon$.

\subsection{Eigenmodes in the limit of vanishing $\boldsymbol{\epsilon}$}

In general, there exist multiple eigenfrequencies,
\begin{equation}
\omega \doteq -\pd_t \theta, 
\end{equation}
that correspond to a local wave vector $\vec{k}$. At vanishingly small $\epsilon$, these eigenfrequencies are found from the GO limit of \Eq{eq:schrodinger}, namely,
\begin{equation}
\omega \xi = H_0(\vec{x},\vec{k}) \xi,
\label{eq:eigen}
\end{equation}
where $H_0(\vec{x},\vec{k}) \doteq \vec{\lambda} \cdot \vec{k}/n$. Notice that the matrix $\mathcal{A}$ does not enter $H_0$, because $\mc{A}$ is of the first order in $\epsilon$. 

In \Eq{eq:eigen}, there exists in general six independent eigenmodes, since $H_0$ is Hermitian. These modes can be readily obtained from
\begin{equation}
\omega^2 \xi = H_0^2 \xi = \frac{1}{n^2} 
		\begin{pmatrix}
			(\vec{\alpha}\cdot \vec{k})^2 	& 0 \\ 0 & (\vec{\alpha}\cdot \vec{k})^2 
		\end{pmatrix}
		\xi.
\end{equation}
Denoting the components of $\xi$ as $\xi^T \doteq \begin{pmatrix}	\vec{a}_1^T & \vec{a}_2^T \end{pmatrix}$, the equation for the first three components is
\begin{align}
\omega^2 \vec{a}_1 
	=& \frac{1}{n^2}  (\vec{\alpha}\cdot \vec{k}) (\vec{\alpha}\cdot \vec{k})  \vec{a}_1 \notag \\
	=& -\frac{1}{n^2}  \vec{k} \times (\vec{k} \times \vec{a}_1) \notag \\
	=& \frac{1}{n^2} \left[  k^2 \vec{a}_1 -(\vec{k}\cdot \vec{a}_1) \vec{k}  \right],
\end{align}
where, in the second line, we have used \Eq{eq:prop_I}. As usual, two eigenmodes are related to the propagation of longitudinal ($\vec{k} \times \vec{a}_1=0$) modes with zero frequency, while the other four eigenmodes correspond to the propagation of transversal ($\vec{k} \cdot \vec{a}_1=0$) EM wave modes in the limit of vanishing $\epsilon$.

We will particularly be interested in transversal EM modes with positive phase velocities, such that
\begin{gather}
\omega = k/n.
\label{eq:omega}
\end{gather}
Corresponding to this frequency, there are two orthonormal eigenvectors $h_q$, which are given by
\begin{align}
h_1(\vec{k}) = \frac{1}{\sqrt{2}}
	\begin{pmatrix}
		\vec{e}_1  \\
		\vec{e}_2 
	\end{pmatrix}, 
& &
h_2(\vec{k}) = \frac{1}{\sqrt{2}}
	\begin{pmatrix}
		\vec{e}_2  \\
		-\vec{e}_1 
	\end{pmatrix},
\label{eq:hq}
\end{align}
where $\vec{e}_1(\vec{k})$ and $\vec{e}_2(\vec{k})$ are any two orthonormal vectors in the plane normal to $\vec{e}_\vec{k} \doteq \vec{k}/k$. A right-hand convention is adopted, such that $\vec{e}_1 \times \vec{e}_2=\vec{e}_\vec{k}$. Obviously, $h_1$ and $h_2$ are eigenvectors of $H_0$. For example,
\begin{align}
H_0 h_1 = & \frac{1}{n} (\vec{k}\cdot \vec{\lambda}) h_1 \notag \\
		= & \frac{i}{n\sqrt{2}}	
					\begin{pmatrix} (\vec{\alpha} \cdot \vec{k}) \vec{e}_2\\ 		-	(\vec{\alpha} \cdot \vec{k}) \vec{e}_1 \end{pmatrix} \notag \\
		= & \frac{  k  }{ n\sqrt{2}}	
					\begin{pmatrix} 	\vec{e}_1  \\  \vec{e}_2  \end{pmatrix} \notag \\
		= & \omega h_1,	
\end{align}
where in the third line, \Eq{eq:prop_I} was used. A similar calculation follows for $h_2$.

\subsection{Eigenmode decomposition}

Since $h_q$ form a complete basis, one can write $\xi = h_q \phi^q$, where $\phi^q$ are scalar functions. We will assume that only transverse modes with positive frequencies are actually excited (we call these modes ``active''), whereas others acquire nonzero amplitudes only through the medium inhomogeneity or the finite width of the EM wave packet (we call these modes ``passive''). In this case,
\begin{equation}
\phi^q = \left\{
\begin{array}{ll}
\mc{O}(\epsilon^0), & q = 1,2,\\
\mc{O}(\epsilon^1), & q \ne 1,2.\\
\end{array}
\right.
\end{equation}
[For future references, note that $\phi^{1,2}(t, \vec{x})$ describe the envelopes corresponding to the linearly polarized modes with the electric field $\vec{E}$ aligned parallel to the unit vectors $\vec{e}_1$ and $\vec{e}_2$, respectively.] For given $\phi^{1,2}$, one can always calculate the amplitudes of passive modes using the complete set of Maxwell's equations, but this will not be necessary for our purposes. As shown in \Ref{Ruiz:2015vq}, due to the mutual orthogonality of all $h_q$, the contribution of passive modes to $\mcc{L}$ is $o(\epsilon)$, so it can be neglected entirely. In other words, for the purpose of calculating $\mcc{L}$, it is sufficient to adopt $\xi \approx h_1 \phi^1 + h_2 \phi^2$, even though the true $\xi$ may have nonzero projections also on other $h_q$. 

It is convenient to rewrite this decomposition in a matrix form,
\begin{equation}
\xi  =  \Xi   \, \phi ,
\label{eq:representation}
\end{equation}
where 
\begin{equation}
\phi (t,\vec{x}) \doteq \begin{pmatrix}  \phi^1 	\\ 		\phi^2    \end{pmatrix},
\end{equation}
and $\Xi$ is a $6\times 2$ matrix having $h_q$ as its columns: \ie
\begin{equation}
\Xi(	\vec{k}	) \doteq \frac{1}{\sqrt{2}} 
			\begin{pmatrix}   \vec{e}_1 	&		\vec{e}_2	\\		
					 			  \vec{e}_2	& 		-\vec{e}_1   
					 			  \end{pmatrix} .
\label{eq:proj}
\end{equation}
[Below, we also consider $\Xi$ as a function of $x^\mu$ in the sense that $\Xi = \Xi(\vec{k}(x^\mu))$.] Then, inserting \Eq{eq:representation} into \Eq{eq:lagr}, one obtains \cite{Ruiz:2015vq}
\begin{equation} \label{eq:lagr_II}
\mcc{L} = \mc{K} - \phi^\dag \left( \mc{E} - \mc{U} \right) \phi +	o(\epsilon)   ,
\end{equation}
where
\begin{gather}
 	\mc{K}  \doteq   \frac{i}{2} \left[ \phi^\dag (d_t \phi) - \cc \right]  , \label{eq:kinetic}  \\
	\mc{E}  \doteq  \pd_t \theta + \frac{k}{n}  ,   \label{eq:energy} \\
  	 \mc{U}  \doteq  \Xi^\dag \mc{M} \Xi + \frac{i}{2}  \left[ \Xi^\dag \gamma^\mu  (\pd_\mu \Xi) - \cc \right].   
\label{eq:potential}
\end{gather}
Here, $d_t \doteq \pd_t + \vec{v}_0 \cdot \del$ is a convective derivative associated to the zeroth order (in $\epsilon$) geometrical optics velocity field,
\begin{equation}
\vec{v}_0 \doteq  \frac{\pd \omega}{\pd \vec{k}} = \frac{1}{ n }\vec{e}_\vec{k}.
\end{equation}
The terms $\mathcal{K}$ and $\mathcal{U}$, which are of order $\epsilon$, represent corrections to the standard, lowest-order GO Lagrangian density. Below, $\mc{U}$ will be calculated explicitly, whereas the higher-order terms, $o(\epsilon)$, will be neglected.

\subsection{Stern-Gerlach Hamiltonian}

Let us search for a tractable expression for $\mc{U}$, which in \Refs{Ruiz:2015uba,Ruiz:2015vq} was called the ``Stern-Gerlach" Hamiltonian. Regarding the first term on the right hand side of \Eq{eq:potential}, a straightforward calculation gives
%
\begin{align}
\Xi^\dag \mc{M} \Xi 
		=&	\frac{1}{2\sqrt{2} n} \Xi^\dag
					 \begin{pmatrix}  0 & \vec{\alpha} \cdot \del	\ln	Z \\  
					 						\vec{\alpha} \cdot \del	\ln	Z  &  0    
					 \end{pmatrix} 
					\begin{pmatrix}   		
								\vec{e}_1  		& 			\vec{e}_2  		\\
								\vec{e}_2  		&			- \vec{e}_1 		
					\end{pmatrix}
		\notag \\
		=&	\frac{i}{4n Z } 
					\begin{pmatrix}
								\vec{e}^1  		&			\vec{e}^2  		 \\
								\vec{e}^2  		&			- \vec{e}^1 		
					\end{pmatrix}
					 \begin{pmatrix}  \del Z 	\times \vec{e}_2 		& 	& -	 \del Z		\times \vec{e}_1		\\  
					 					 \del Z		\times \vec{e}_1  	&  &		 \del Z		\times \vec{e}_2    
					 \end{pmatrix} 
		\notag \\
		=& \, 0.
\end{align}
%
%
Here, we introduced the notation $\vec{e}^1	\doteq 	\vec{e}_1^T$ and $\vec{e}^2	\doteq 	\vec{e}_2^T$. Notably, these can be understood as adjoint basis vectors; \eg $\vec{e}^2 \vec{e}_1\equiv (\vec{e}^2 )_i (\vec{e}_1 )^i \equiv \vec{e}_2 \cdot \vec{e}_1=0$. Hence, \Eq{eq:potential} is simplified down to
\begin{equation}
\mc{U} =  \frac{i}{2}  \left[ \Xi^\dag \gamma^\mu  (\pd_\mu \Xi) - \hc \right] .
\end{equation}
Using that $\pd_\mu \Xi = (\pd \Xi/\pd k_j)(\pd_\mu k_j)$, one can rewrite $\mc{U}$ as follows [18]:
\begin{equation}
\mc{U} = 
			\frac{i}{2} \dot{\vec{k}} \cdot  \left[  \Xi^\dag (\pd_{\vec{k}} \Xi) - (\pd_{\vec{k}} \Xi^\dag) \Xi   \right] .
\label{eq:Ured}
\end{equation}
Here, $\dot{\vec{k}}(t)$ is given by the lowest-order (in $\epsilon$) ray equation,
\begin{equation}
\dot{\vec{k}}(t) = -\pd_\vec{x}\omega(\vec{x},\vec{k}) , \label{eq:ray}
\end{equation}
and $\omega$ is given by \Eq{eq:omega}. Using \Eq{eq:proj}, we obtain
\begin{multline}
\Xi^\dag \frac{\pd \Xi  }{\pd k^j} - \frac{\pd \Xi^\dag }{\pd k^j}  \Xi =\\
			\begin{pmatrix}
				0									&						\vec{e}^1 	\displaystyle\frac{\pd}{\pd k^j}  \vec{e}_2		- \vec{e}^2 	\displaystyle\frac{\pd}{\pd k^j} \vec{e}_1		\notag \\
				\vec{e}^2  	\displaystyle\frac{\pd}{\pd k^j} \vec{e}_1	 	-  \vec{e}^1 	\displaystyle\frac{\pd}{\pd k^j}  \vec{e}_2		& 0 
			\end{pmatrix}.
\label{eq:U_aux}
\end{multline}
Since $\vec{e}^1 \vec{e}_2 =0$, \Eq{eq:Ured} leads to
\begin{equation}
\mc{U} = -\dot{\vec{k}}  \cdot \vec{A}(\vec{k}) \sigma_y , 
\label{eq:SGpotential}
\end{equation}
where $\sigma_y$ is the $y$-component of the Pauli matrices,
\begin{equation}
\sigma_y=
	\begin{pmatrix}
		0 	& 	-i \\
		i	&	0
	\end{pmatrix},
\end{equation}
and $\vec{A} (\vec{k})$ is a vector with components given by
\begin{equation}
A_j	(\vec{k}) \doteq \vec{e}^1 \displaystyle\frac{\pd}{\pd k^j} \vec{e}_2	.
\end{equation}
For example, one may choose \cite{foot:orthonormal}
\begin{align}
\vec{e}_1(\vec{k}) \doteq 	
		\begin{pmatrix}  \frac{k_x k_z }{k \sqrt{ k_x^2+k_y^2 } } \\ \frac{k_y k_z}{ k \sqrt{ k_x^2+k_y^2 } } \\ - \frac{ \sqrt{ k_x^2+k_y^2 } }{ k } \end{pmatrix} , 
& &
\vec{e}_2(\vec{k}) \doteq 	
		\begin{pmatrix}  -\frac{k_y }{  \sqrt{ k_x^2+k_y^2 } } \\ \frac{k_x }{  \sqrt{ k_x^2+k_y^2 } } \\ 0 \end{pmatrix},
\label{eq:evec}
\end{align}
so that
\begin{equation}
\vec{A} (\vec{k}) = \frac{\vec{e}_\perp \times \vec{e}_\vec{k} }{ k_\perp  },
\end{equation}
where $\vec{k}_\perp^T \doteq \begin{pmatrix}		k_x, k_y, 0 \end{pmatrix}$, and $\vec{e}_\perp \doteq \vec{k}_\perp/k_\perp$; or, more explicitly,
\begin{equation}
\vec{A} (\vec{k}) = \frac{k_z}{kk_\perp^2}
	\begin{pmatrix}
			 k_y		\\ 			
			-k_x 		\\ 		
			0 
	\end{pmatrix}.		
\label{eq:def_A}
\end{equation}
%

\subsection{Lagrangian density: summary}

Substituting \Eq{eq:SGpotential} into \Eq{eq:lagr_II}, we can express the Lagrangian density as
\begin{multline}
\mcc{L} = -\phi^\dag \left( \pd_t \theta + \frac{k}{n} \right)	\phi	\\
			+  \frac{i}{2} \left(	\phi^\dag d_t \phi - \cc \right) 			
			- \phi^\dag  \sigma_y  \Sigma  \phi,
\label{eq:lagr_reduced}
\end{multline}
where
\begin{equation}
\Sigma		\doteq 	\dot{\vec{k}}\cdot \vec{A}(\vec{k}) .
\label{eq:Sigma}
\end{equation}
Let us also use a variable transformation,
\begin{equation}
\phi(t,\vec{x})= \, Q \, \eta(t,\vec{x}), 
\end{equation}
where
\begin{equation}
	Q \doteq \frac{1}{\sqrt{2}}
			\begin{pmatrix}
				1	&  1	\\		
				i	&	-i
			\end{pmatrix} ,			
\end{equation}
and $\eta(t,\vec{x})$ is a new vector with components denoted as
\begin{equation}
\eta(t,\vec{x})  \doteq 			
			\begin{pmatrix}
				\eta_+ \\
				\eta_-
			\end{pmatrix}.
\end{equation}
Hence, the Lagrangian density \eq{eq:lagr_reduced} can be expressed as
\begin{multline}
	\mcc{L} = -\eta^\dag \left( \pd_t \theta + k/n \right) \eta  \\
			+  \frac{i}{2} \left(	\eta^\dag d_t \eta - \cc \right) 			
			- \eta^\dag  \sigma_z  \Sigma \eta,
\label{eq:lagr_diag}
\end{multline}
where $\sigma_z$ is another Pauli matrix,
\begin{equation}
\sigma_z=
	\begin{pmatrix}
		1		&		0	\\
		0		&		-1		
	\end{pmatrix}.
\end{equation}
Here, $\eta_\pm(t,\vec{x}) $ describe envelopes corresponding to right-hand and left-hand circularly polarized modes, respectively (as defined from the point of view of the source). 

The first line of the right hand side of \Eq{eq:lagr_diag} represents the lowest order GO Lagrangian density for these modes. Additionally, the second line of \Eq{eq:lagr_diag}, which contains $\mc{O}(\epsilon)$ terms, introduces polarization effects, as will be explained below.

\section{Continuous wave model}
\label{sec:wave}

Substituting \Eqs{eq:omega} and \eq{eq:ray} in \Eq{eq:Sigma}, we can rewrite $\mcc{L}$ as
\begin{multline}
\mcc{L} = - \eta^\dag \left(\pd_t \theta + \frac{k}{n}\right) \eta \\
				+ \frac{i}{2}\,(\eta^\dag d_t \eta - \cc ) - \eta^\dag \sigma_z 	\Sigma(\vec{x}, \vec{k}) \eta,
\label{eq:lagr_reduced_II}
\end{multline}
where
\begin{equation}
\Sigma(\vec{x}, \vec{k}) = \frac{k}{n^2}  \vec{A}(\vec{k}) \cdot \del n.
\label{eq:Sigma_II}
\end{equation}
In this form, the Lagrangian density is analogous to that of a semiclassical Pauli particle and thus can be approached similarly \cite{Ruiz:2015uba}. Let us adopt the representation $\eta = z\sqrt{\mc{I}}$, where $\mc{I}(t,\vec{x}) \doteq \psi^\dag \psi$ is the action density, and $z(t,\vec{x})$ is a unit polarization vector ($z^\dag z \equiv 1$). Since the common phase of the two components of $z$ can be attributed to $\theta$, we can parameterize $z$ in terms of just two real functions, $\vartheta(t,\vec{x})$ and $\zeta(t,\vec{x})$:
\begin{gather}\notag
z(\vartheta, \zeta) = 
\begin{pmatrix}
e^{i\vartheta} \cos (\zeta/2) \\
e^{-i\vartheta} \sin (\zeta/2)
\end{pmatrix}.
\end{gather}
Like in the case of the Pauli particle, $\zeta$ determines the relative fraction of ``spin-up'' and ``spin-down'' quanta, \ie those corresponding to left-hand and right-hand polarizations. Also, one can understand $\vec{S} \doteq z^\dag \vec{\sigma} z/2$ as the wave ``spin vector'' ($\vec{\sigma}$ denotes the three Pauli matrices, as usual \cite{Ruiz:2015uba, Ruiz:2015vq}), or, up to a constant factor, as the Stokes vector \cite{Kravtsov:2007cl, Kravtsov:2010te}.

Using the above parameterization, the Lagrangian density is rewritten as
\begin{gather}
\mcc{L} = - \mc{I} \left[\pd_t \theta + \frac{k}{n} + (d_t \vartheta + \Sigma )\cos \zeta \right],
\end{gather}
which leads to four ELEs. The first one is the action conservation theorem,
\begin{gather}\label{eq:act}
\delta \theta : \quad \pd_t \mc{I} + \del \cdot ( \mc{I} \vec{V}) = 0.
\end{gather}
The flow velocity is given by $\vec{V} = \vec{v}_0 + \vec{u}$, and
\begin{gather}
\vec{u} \doteq 
		\frac{\pd}{\pd\vec{k}}\left(\frac{\vec{k} \cdot \del \vartheta}{n k} + \Sigma\right) \cos \zeta.
\end{gather}
Notice that $\vec{u}$ represents the polarization-driven deflection of the ray's ``center of gravity" predicted in \Refs{Dugin:1991tx,Dooghin:1992kr,Liberman:1992bz}. The second ELE is a Hamilton-Jacobi equation, 
\begin{gather}
\delta \mc{I} : \quad  \pd_t \theta + \frac{k}{n} + (d_t \vartheta + \Sigma) \cos \zeta = 0,
\end{gather}
whose gradient yields an equation for $\pd_t \vec{k}$, \ie the momentum equation \cite{ Ruiz:2015uba}. The third ELE is
\begin{gather}\label{eq:zeta}
\delta \zeta : \quad  d_t \vartheta = -\Sigma.
\end{gather}
As it will become clear below, this describes the rotation of the wave polarization. Finally, the fourth ELE is
\begin{gather}\label{eq:vartheta}
\delta \vartheta: \quad  \pd_t (\mc{I} \cos \zeta) + \del \cdot ( \mc{I} \vec{v}_0 \cos \zeta) = 0.
\end{gather}

Together, \Eqs{eq:act}-\eq{eq:vartheta} provide a complete ``fluid'' description of continuous waves. Note that waves are allowed to be nonstationary and ``entangled'', \ie contain mixed polarization. In fact, a combination of \Eq{eq:vartheta} with \Eq{eq:act} gives
\begin{equation}
\pd_t (\cos \zeta ) +  \vec{V}\cdot \del  (\cos \zeta) = \frac{1}{\mc{I}} \del \cdot \left( \mc{I} \vec{u} \cos \zeta \right),
\end{equation}
which shows that $\zeta$ is generally not conserved; \ie the numbers of ``spin-up'' and ``spin-down'' quanta necessarily oscillate, unless a wave is homogeneous. This can be interpreted as an effective \textit{zitterbewegung} of a classical wave in an inhomogeneous medium.

\section{Ray dynamics: canonical representation}
\label{sec:ray_can}

\subsection{Basic equations}
\label{sec:ray_canonical}

The ray equations corresponding to the above field equations can be obtained as a point-particle limit. In this limit, $\mc{I}$ can be approximated with a delta function
\begin{gather}
\mc{I}(t, \vec{x}) = \delta(\vec{x} - \vec{X}(t)),
\end{gather}
where $\vec{X}$ is the location of the wave packet. As in \Refs{Ruiz:2015uba,Ruiz:2015vq}, the Lagrangian density \eq{eq:lagr_diag} yields a point-particle Lagrangian, $L	\doteq \int \mcc{L}~ \mathrm{d}^3x$, specifically,
\begin{gather}
L = \vec{P} \cdot \dot{\vec{X}} - \frac{c P}{n} + \frac{i}{2}\,(Z^\dag\dot{Z} - \dot{Z}^\dag Z) - \Sigma(\vec{X}, \vec{P}) \,Z^\dag \sigma_z Z. \label{eq:lagr_point}
\end{gather}
Here $\vec{P}( t ) \doteq \del \theta( t , \vec{X}(t) )$ is the canonical momentum, $P=|\vec{P}|$, and $Z(t)\doteq z(t,\vec{X}(t))$ is a complex scalar function. The refraction index and $\Sigma$ are evaluated at $\vec{x}=\vec{X}$ and $\vec{k}=\vec{P}$; \eg $n=n(\vec{X}(t))$. Also, the speed of light constant, $c$, has been reintroduced for clarity.

Treating $\vec{X}(t)$, $\vec{P}(t)$, $Z(t)$, and $Z^\dag(t)$ as independent variables, we obtain the following ELEs:
\begin{align}
\delta \vec{P}: \quad 	&	\dot{\vec{X}}(t)		= \frac{c\vec{P}}{nP}
										+  (\pd_\vec{P}   \Sigma) \,Z^\dag \sigma_z Z ,
									\label{eq:X}   \\
\delta \vec{X}: \quad 	&	\dot{\vec{P}}(t)		= \frac{cP}{n^2} \del n 
										- (\pd_\vec{X}  \Sigma) \,Z^\dag \sigma_z Z ,
									\label{eq:P}   \\
\delta Z^\dag: \quad 	&	\dot{Z}(t)= -i \Sigma\sigma_z Z ,
									\label{eq:Z}  \\
\delta Z: \quad 			&	\dot{Z}^\dag(t)= i \Sigma Z^\dag \sigma_z . 
									\label{eq:Z_dag} 
\end{align}
Together with \Eqs{eq:def_A} and \eq{eq:Sigma_II}, \Eqs{eq:X}-\eq{eq:Z_dag} form a complete set of equations. The first terms on the right hand side of \Eqs{eq:X} and \eq{eq:P} describe the ray dynamics in the GO limit. The second terms describe the coupling of the mode polarization and the ray curvature.

\subsection{Polarization dynamics}

To better understand the polarization equations, let us rewrite \Eq{eq:Z} as an equation in the basis of linearly polarized modes, \ie for $\Phi(t) \doteq \phi(t,\vec{X}(t))$:
\begin{gather}
\dot{\Phi} = Q \dot{Z} = -i\Sigma Q \sigma_z Z = -i\Sigma (Q \sigma_z Q^{-1}) \Phi = -i\Sigma \sigma_y \Phi.
\end{gather}
[This equation could also be obtained if the ray equations were derived directly from the Lagrangian density \eq{eq:lagr_reduced}.] Since $\Sigma$ is a scalar, and $\sigma_y$ is constant, this can be readily integrated, yielding \cite{foot:exponent}
\begin{gather}
\Phi = \exp( -i \Theta \sigma_y) \Phi_0 = (\mathbb{I}_2 \cos\Theta - i \sigma_y \sin \Theta)\Phi_0,
\end{gather}
where $\Theta(t) \doteq \int^t_0 \Sigma(t')\,dt'$ can be recognized as the wave Berry phase \cite{Berry:1984jv}, $\Sigma(t) \doteq \Sigma(\vec{X}(t), \vec{P}(t))$ and $\Phi_0$ is a vector determined by initial conditions. This result can be also be expressed explicitly as follows:
\begin{gather}
\Phi = \left(
\begin{array}{cc}
\cos\Theta & -\sin\Theta \\
\sin\Theta & \cos\Theta \\
\end{array}
\right)\Phi_0.
\end{gather}
It is seen then that the polarization of the EM field rotates at the rate $\Sigma(t)$ in the reference frame defined by the basis vectors $(\vec{e}_1, \vec{e}_2)$. Such rotation of the polarization plane, also known as \textit{Rytov rotation}, was studied theoretically in \Refs{Rytov:1938wp,Vladimirskii:1941wa,Berry:1987cj} and observed experimentally in \Ref{Tomita:1986jo}. Clearly, \Eq{eq:zeta} describes the same effect.

\subsection{Ray dynamics for pure states}
\label{sec:ray_pure}

If a ray corresponds to a strictly circular polarization, such that $\sigma_z z = \pm z$, the Lagrangian \eq{eq:lagr_point} can be simplified down to
\begin{gather}
L_\pm = \vec{P} \cdot \dot{\vec{X}} - \frac{cP}{n} \mp \Sigma(\vec{X}, \vec{P}),
\label{eq:lagr:pure}
\end{gather}
where $L_\pm$ governs the propagation of right and left polarization modes, respectively. This Lagrangian has a canonical form, $L_\pm=\vec{P}\cdot \dot{\vec{X}}- H_\pm(\vec{X},\vec{P})$, where $H_\pm$ is a Hamiltonian given by
\begin{equation}
H_\pm(\vec{X},\vec{P}) = \frac{cP}{n(\vec{X}) } \pm \frac{cP}{n^2(\vec{X}) } \vec{A}(\vec{P}) \cdot \del n(\vec{X} ).
\label{eq:ham_canonical}
\end{equation}
The variables $\vec{X}$ and $\vec{P}$ serve as the canonical coordinate and momentum, so they satisfy canonical Hamilton's equations,
\begin{align}
\delta \vec{P}: \quad 	&	\dot{\vec{X}}(t)		= \frac{c\vec{P}}{nP}
										\pm  \pd_\vec{P}   \Sigma  ,
									\label{eq:X_II}   \\
\delta \vec{X}: \quad 	&	\dot{\vec{P}}(t)		= \frac{cP}{n^2} \del n 
										\mp \pd_\vec{X}  \Sigma  .
									\label{eq:P_II}   
\end{align}
Since $H_\pm$ is time-independent, one also readily obtains energy (frequency) conservation, 
\begin{equation}
H_\pm(\vec{X}(t),\vec{P}(t))=\mathrm{const},
\end{equation}
along the ray trajectory.

\begin{figure*}
  	\includegraphics[width=\textwidth]{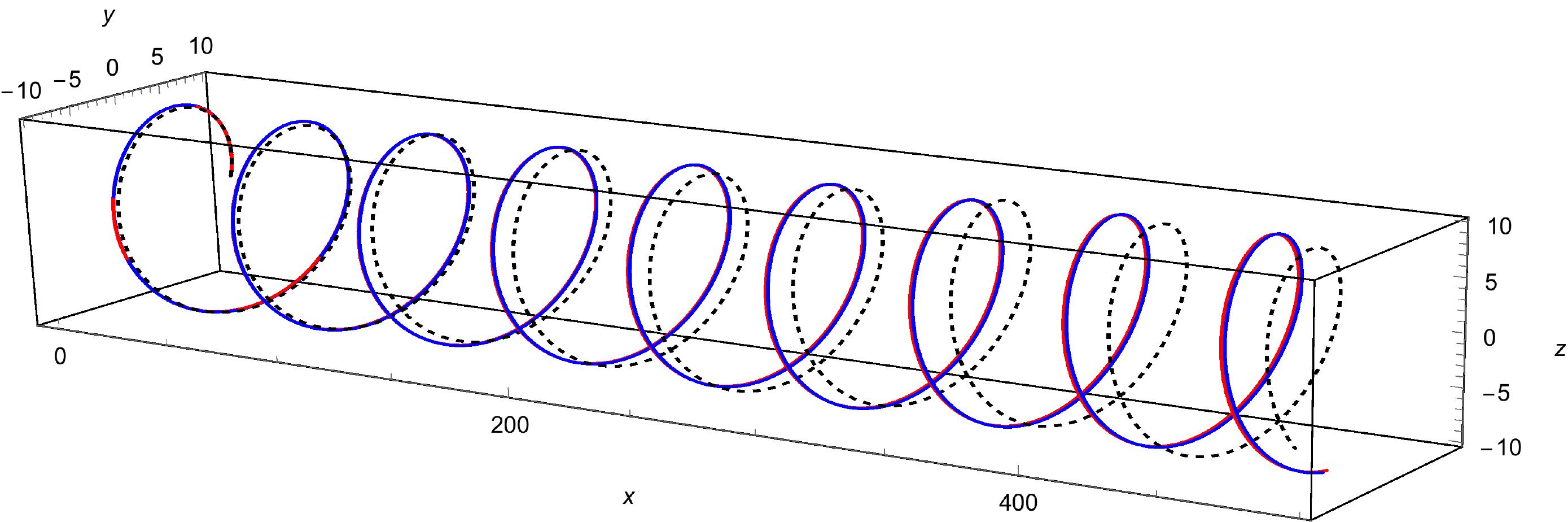}
  	\caption{
  	Comparison between ray trajectories predicted by the canonical Lagrangian \eq{eq:lagr:pure} and the non-canonical Lagrangian  \eq{eq:lagr:bliokh}. The red line represents $\vec{X}(t) - \vec{A}(\vec{P}(t))$, where $\vec{X}(t)$ and $\vec{P}(t)$ are found by numerical integration of \Eqs{eq:X_II} and \eq{eq:P_II}. The blue line shows $\vec{x}(t)$, as found by numerical integration of \Eqs{eq:x_II} and \eq{eq:p_II}. The black dashed line plots the ``spinless'' ray trajectory governed by the lowest-order GO ray Lagrangian $L_0$ given by \Eq{eq:lagr_no_spin}. The refraction index is chosen to be $n(\vec{x})=1+\exp[-(y^2+z^2)/\ell^2]$, the characteristic width is $\ell=10$, and $\vec{A}$ is chosen in the form \eq{eq:def_A}. The length unit is $a\doteq \sqrt{2}/p_0$. The initial location is $\vec{x}_0=(0,10,0)$, and the initial momentum is $\vec{p}_0=(1,0,1)/a$, so $\epsilon\sim 1/(p_0 \ell)\sim 0.07$. It is seen that the canonical Lagrangian and the noncanonical Lagrangian predict results that, for the specified parameters, are essentially indistinguishable from each other yet differ noticeably from those yielded by $L_0$. }
  	\label{fig:figure}
\end{figure*}

\section{Ray dynamics: non-canonical representation}
\label{sec:ray_noncan}

\subsection{Ray variables}
\label{sec:ray_variables}

If the point-particle limit is taken without explicitly invoking \Eq{eq:Sigma}, an alternative representation of the ray Lagrangian can be obtained that connects our results to those found in existing literature. For pure states, this alternative Lagrangian is given by
\begin{equation}
L_\pm = \vec{p} \cdot \dot{\vec{x}} - \frac{cp}{n(\vec{x}) }  \mp  \dot{\vec{p}} \cdot \vec{A}(\vec{p}) ,
\label{eq:lagr:bliokh}
\end{equation}
where $\dot{\vec{p}} \cdot \vec{A}(\vec{p})$ is known as the ``Berry connection'' term \cite{Bliokh:2008km}. Notice, in particular, that adding $\pd_\vec{k}\chi(\vec{k})$ to $\vec{A}(\vec{k})$, where $\chi(\vec{k})$ is an arbitrary scalar function, changes $L_\pm$ merely by a perfect time derivative and thus does not affect the motion equations. Also notice that we have introduced a different notation for the ray variables [$(\vec{x}, \vec{p})$ instead of $(\vec{X}, \vec{P})$] for the following reason. 

On one hand, as an approximation of the Lagrangian \eq{eq:lagr:pure}, \Eq{eq:lagr:bliokh} is expected to yield dynamics similar to that yielded by \Eq{eq:lagr:pure}. On the other hand, notice that \Eq{eq:lagr:bliokh} does not have a canonical form. This means that $(\vec{x}, \vec{p})$ do not satisfy Hamilton's equations and thus, clearly, cannot be the same as $(\vec{X}, \vec{P})$. To understand the connection between the two sets of variables, let us rewrite \Eq{eq:lagr:bliokh} as
\begin{equation}
L_\pm = \vec{p} \cdot \dot{\vec{x}} - \frac{cp}{n}  \mp  \frac{d}{d t}\left( \vec{p} \cdot \vec{A}\right) \pm \vec{p} \cdot \frac{d \vec{A}}{dt} .
\end{equation}
Dropping the perfect time derivative and introducing
\begin{equation}
\vec{q}\doteq \vec{x} \pm \vec{A},
\end{equation}
we obtain a Lagrangian in a canonical form:
\begin{equation}
L_\pm = \vec{p} \cdot \dot{\vec{q}} - \frac{cp}{n(\vec{q}\mp \vec{A} (\vec{p})  )}.
\end{equation}
The quantity $cp/n(\vec{x})$ serves as the canonical energy and is conserved,
\begin{equation}
cp/n(\vec{x}) = \text{const}.
\label{eq:energy_conser}
\end{equation}
Also, since $1/n(\vec{x})$ is assumed smooth, one can replace this function with its first-order Taylor expansion. Then,
\begin{gather}
L_\pm = \vec{p} \cdot \dot{\vec{q}} - \frac{cp}{n(\vec{q}) } 
		\mp \frac{cp}{n^2(\vec{q})} \vec{A}(\vec{p}) \cdot \del n(\vec{q} ),
\end{gather}
where we omitted terms $\mc{O}(\epsilon^2)$, as usual. By comparing this with \Eq{eq:lagr:pure}, we find that
\begin{align}
\vec{x} = \vec{X} \mp \vec{A}(\vec{P}),	 &		&
\vec{p} = \vec{P}. \label{eq:con_x}
\end{align}
Notice that $|\vec{x} - \vec{X}|$ is of the order of the wavelength, i.e., small enough to make $\vec{x}$ and $\vec{X}$ equally physical as measures of the ray location.

\subsection{Ray equations}

The equations for $\vec{x}$ and $\vec{p}$ can be obtained by combining \Eqs{eq:X_II}, \eq{eq:P_II}, and \eq{eq:con_x}, or they can be derived directly as ELEs corresponding to the Lagrangian \eq{eq:lagr:bliokh}:
\begin{align}
\delta \vec{p}: 	&  \quad  \dot{\vec{x}}(t) 	=  \frac{c\vec{p}}{np}   \pm  \dot{\vec{p}} \times  \left( \del_\vec{p} \times \vec{A} \right)    , \label{eq:x}  \\
\delta \vec{x}: 	& 	\quad \dot{\vec{p}}(t)	= \frac{cp}{n^2}  \del n. 		\label{eq:p} 
\end{align}
Using \Eq{eq:def_A}, we can also rewrite them as
\begin{align}
\dot{\vec{x}}(t) =&   \frac{c\vec{p}}{np}     \pm   \frac{  \dot{\vec{p} } \times \vec{p}   }{p^3} , 		\label{eq:x_II}\\
\dot{\vec{p}}(t)= &   \frac{cp}{n^2}  \del n. \label{eq:p_II}		
\end{align}

Equations \eq{eq:x_II} and \eq{eq:p_II} were reported previously in \Refs{Onoda:2004ij,Onoda:2006gg}. The equations presented in  \Refs{Bliokh:2004ja,Bliokh:2004kp,Bliokh:2005be,Bliokh:2007fr,Bliokh:2008km} can also be obtained from \Eqs{eq:x_II} and \eq{eq:p_II} for the rescaled momentum $\vec{\rho} \doteq n\vec{p}/(cp)$ and rescaled ``time" $ds \doteq c\,dt/n(\vec{x}(t))$. Specifically, one gets
\begin{gather}
\vec{x}' = \frac{\vec{\rho}}{\rho}  \pm \lambdabar\,
\frac{\vec{\rho}' \times \vec{\rho}}{\rho^3}, 
\quad \vec{\rho}' = \del n,
\end{gather}
where primes denote derivatives with respect to $s$, and we used that $\lambdabar \doteq n(\vec{x})/p$ is a constant of motion, as seen from \Eq{eq:energy_conser}.

\subsection{Comparison of the two models}

As we showed explicitly in \Sec{sec:ray_variables} (and also by construction), the canonical Lagrangian \eq{eq:lagr:pure} and the noncanonical Lagrangian \eq{eq:lagr:bliokh} differ only by $\mc{O}(\epsilon^2)$ and thus are equivalent within their accuracy domain. (The same applies to the earlier theories \cite{Onoda:2004ij,Onoda:2006gg,Bliokh:2004ja,Bliokh:2004kp,Bliokh:2005be,Bliokh:2007fr,Bliokh:2008km} too, where the ray Lagrangians are also derived to the first order in $\epsilon$.) This means, in particular, that the effect of the Berry connection term in \Eq{eq:lagr:bliokh} can be attributed simply to the choice of coordinates, while the underlying physics can also be described by the canonical Lagrangian \eq{eq:ham_canonical}. 

The advantage of the Lagrangian \eq{eq:lagr:pure} is its manifestly symplectic structure, which is convenient, for instance, for numerical simulations \cite{Bridges:2001aa}. On the other hand, the Lagrangian \eq{eq:lagr:bliokh} leads to ``gauge-invariant'' equations, in the sense that they are manifestly independent of the coordinate representation for the basis vectors, $\vec{e}_1$ and $\vec{e}_2$. Also, assuming that $\vec{A}$ is chosen in the form \eq{eq:def_A}, using the non-canonical Lagrangian \eq{eq:lagr:bliokh} can be advantageous when $p_x^2 + p_y^2 \to 0$, because the right-hand side in \Eqs{eq:x_II} and \eq{eq:p_II} remains finite in this case unless $p \to 0$. Therefore, whether the canonical or noncanonical form is more convenient in a given case depends on the specific application.

To illustrate how accurate the agreement is between the two models, we also performed comparative numerical simulations. Figure 1 shows the ray trajectories for a right polarized wave using the canonical \eq{eq:lagr:pure} and the non-canonical \eq{eq:lagr:bliokh} representations. For completeness, we also show the calculated ray trajectory as determined by the lowest-order GO ray Lagrangian,
\begin{gather}
L_0 = \vec{p} \cdot \dot{\vec{x}} - cp/n(\vec{x}),
\label{eq:lagr_no_spin}
\end{gather}
which does not account for polarization effects. As anticipated, the ray trajectories predicted by the Lagrangians \eq{eq:lagr:pure} and \eq{eq:lagr:bliokh} are almost identical and yet differ noticeably from the ``spinless'' ray trajectory predicted by \Eq{eq:lagr_no_spin}, namely,
\begin{align}
\delta \vec{p}: 	&  \quad  \dot{\vec{x}}(t) 	=  \frac{c\vec{p}}{np}  ,   \\
\delta \vec{x}: 	& 	\quad \dot{\vec{p}}(t)	= \frac{cp}{n^2}  \del n. 	
\end{align}


\section{Conclusions}
\label{sec:conclusion}

In this paper, we study the propagation of electromagnetic waves in isotropic dielectric media with local dispersion under the assumption of small but nonvanishing $\epsilon \doteq \lambda/\ell$, where $\lambda$ is the wavelength, and $\ell$ is the characteristic inhomogeneity scale. It is commonly known that, due to nonzero $\epsilon$, such waves can experience polarization-driven bending of ray trajectories and polarization dynamics that can be interpreted as the precession of the wave ``spin''. Here, we report how Lagrangians describing these effects can be deduced, rather than guessed, within a strictly classical theory. In addition to the commonly known ray Lagrangian that features the Berry connection, a simple alternative Lagrangian is also proposed that naturally has a canonical form. We explain how the two Lagrangians are related and demonstrate their equivalence numerically.  The presented theory captures not only the eigenray dynamics but also the dynamics of continuous wave fields and rays with mixed polarization, or ``entangled'' waves. Our calculation assumes stationary media with isotropic local dispersion, but generalizations to other media are straightforward to do, as will be reported separately.

This work was supported by the NNSA SSAA Program through DoE Research Grant No. DE274-FG52-08NA28553, by the U.S. DoE through Contract No. DE-AC02-09CH11466, by the U.S. DTRA through Research Grant No. HDTRA1-11-1-0037, and by DoD, Air Force Office of Scientific Research, National Defense Science and Engineering Graduate (NDSEG) Fellowship, 32-CFR-168a.



\begin{thebibliography}{35}
\bibitem[{foo({\natexlab{b}})}]{foot:GellMann}
\bibinfo{note}{It is to be noted that these $\vec{\alpha}$ matrices are related
  to the Gell-Mann matrices, which serve as infinitesimal generators of the
  special unitary group $SU(3)$.}
\bibitem[{foo({\natexlab{c}})}]{foot:orthonormal}
\bibinfo{note}{Note that $\vec{e}_1$ and $\vec{e}_2$ correspond to the
  spherical unit vectors $\vec{e}_\theta$ and $\vec{e}_\varphi$ in
  $\vec{k}$-space, respectively.}
\bibitem[{foo({\natexlab{d}})}]{foot:exponent}
\bibinfo{note}{Here, we used the well known Euler formula for Pauli matrices,
  $e^{i a (\vec{n}\cdot \vec{\sigma}) }=\mathbb{I}_2 \cos a+ i (\vec{n}\cdot
  \vec{\sigma}) \sin a$.}
\end{thebibliography}
\end{document}